\documentclass[hidelinks]{llncs}

\usepackage[]{geometry}
\usepackage[english]{babel}
\usepackage[T1]{fontenc}
\usepackage[utf8]{inputenc}
\usepackage{array}
\usepackage[htt]{hyphenat}
\usepackage[makeroom]{cancel}
\usepackage[nolist]{acronym}
\usepackage{algorithm}
\usepackage{algpseudocode}
\usepackage{amsmath}
\usepackage{amssymb}
\usepackage{booktabs}
\usepackage{csquotes}
\usepackage{dashbox}
\usepackage{dtk-logos}
\usepackage{float}
\usepackage{latexsym}
\usepackage{lmodern}
\usepackage{longtable}
\usepackage{mathrsfs}
\usepackage{mathtools}
\usepackage{listings}
\usepackage{multirow}
\usepackage{needspace}
\usepackage{perpage}
\usepackage{pgfplots}
\usepackage{pifont}
\usepackage{setspace}
\usepackage{subcaption}
\usepackage{tikz}
\usepackage{pgf-umlcd}
\usepackage{graphicx}
\usepackage{xcolor}
\usepackage{cite}
\usepackage[pdftitle={muRelBench: MicroBenchmarks for Zonotope Domains},
            pdfauthor={Kenny Ballou and Elena Sherman},
            pdfkeywords={Microbenchmarks, Relational Abstract Domains},
            colorlinks=false]{hyperref}
\usepackage{hyperxmp}
\usepackage[automake,acronym,index,nogroupskip,symbols]{glossaries-extra}

\algnewcommand\algorithmicswitch{\textbf{switch}}
\algnewcommand\algorithmiccase{\textbf{case}}
\algnewcommand\algorithmicassert{\texttt{assert}}
\algnewcommand\Assert[1]{\State{\algorithmicassert(#1)}}%
\algdef{SE}[SWITCH]{Switch}{EndSwitch}[1]{\algorithmicswitch\ #1\ \algorithmicdo}{\algorithmicend\ \algorithmicswitch}%
\algdef{SE}[CASE]{Case}{EndCase}[1]{\algorithmiccase\ #1}{\algorithmicend\ \algorithmiccase}%
\algtext*{EndSwitch}%
\algtext*{EndCase}%

\definecolor{redbg0}{HTML}{FFD8D5}
\definecolor{redbg1}{HTML}{F3B5AF}
\definecolor{greenbg0}{HTML}{C1F2D1}
\definecolor{greenbg1}{HTML}{AEE5BE}

\usetikzlibrary{%
  backgrounds,
  external,
  trees,
  calc,
  external,
  positioning,
  shapes.geometric,
  shapes.misc,
  shapes.symbols,
  shadows,
  arrows,
  arrows.meta,
  fit,
  matrix,
  mindmap,
}

\newcolumntype{?}{!{\vrule width 1.25pt}}

\setabbreviationstyle[acronym]{long-short}

\glsdisablehyper{}

\MakePerPage{footnote}

\makeglossaries[]

\begin{document}

\title{muRelBench: MicroBenchmarking for Zonotope Domains}

\author{%
  Kenny Ballou \orcidID{0000-0002-6032-474X} \inst{1}\and
  Elena Sherman \orcidID{0000-0003-4522-9725} \inst{2}
}
\institute{%
  California State University San Marcos,
  \email{kballou@csusm.edu}
\and
  Boise State University
  \email{elenasherman@boisestate.edu}
}
\authorrunning{Ballou, et~al.}

\begin{acronym}
  \newacronym{AI}{AI}{abstract interpretation}
  \newacronym{CFG}{CFG}{control flow graph}
  \newacronym{DFA}{DFA}{data-flow analysis}
  \newacronym{DFS}{DFS}{depth-first search}
  \newacronym[longplural={difference bounded matrices}]{DBM}{DBM}{difference bounded matrix}
  \newacronym{JVM}{JVM}{Java Virtual Machine}
  \newacronym{LIA}{LIA}{linear integer arithmetic}
  \newacronym{NIA}{NIA}{non-linear integer arithmetic}
  \newacronym{TVPI}{TVPI}{two variables per inequality}
  \newacronym{uTVPI}{uTVPI}{unit two variables per inequality}
  \newacronym{SMT}{SMT}{satisfiability modulo theories}
  \newacronym{SMT-LIB}{SMT-LIB}{satisfiability modulo theories library}
  \newacronym{StInG}{StInG}{Stanford Invariant Generator}

  \newacronym{FS}{FS}{full state}
  \newacronym{FG}{FG}{Full Graph}
  \newacronym{CC}{CC}{Connect Components}
  \newacronym{NN}{NN}{Node Neighbors}
  \newacronym{MN}{MN}{Minimal Neighbors}
  \newacronym{JIT}{JIT}{Just-in-Time}
  \newacronym{GC}{GC}{Garbage Collection}
  \newacronym{UML}{UML}{Unified Modeling Language}
  \newacronym{LUB}{LUB}{Least-Upper Bound}
  \newacronym{CSV}{CSV}{Comma-Separated Values}
\end{acronym}



\maketitle{}

\begin{abstract}

We present \texttt{muRelBench}, a framework for synthetic benchmarks for
weakly-relational abstract domains and their operations.  This extensible
microbenchmarking framework enables researchers to experimentally evaluate
proposed algorithms for numerical abstract domains, such as closure,
least-upper bound, and forget, enabling them to quickly prototype and validate
performance improvements before considering more intensive experimentation.
Additionally, the framework provides mechanisms for checking correctness
properties for each of the benchmarks to ensure correctness within the
synthetic benchmarks.

\end{abstract}



\section{Introduction}\label{sec:intro}%

Zonotopes~\cite{ghorbal-2009-zonoto-abstr}, relational numerical abstract
domains, are widely used in program and system verification using static
analysis and model-checking techniques and, recently, found their way into the
verification of neural networks~\cite{jordan-2022-zontope-neural}.  To reason
about their computations, verifiers manipulate abstract domains through a
predefined set of operations, e.g., \gls{LUB}, closure, or forget
operators~\cite{mine-2006-octag-abstr-domain}.  Such manipulations of abstract
states commonly dominate the computation time of a verifier.  Consequently,
there has been extensive research on improving the efficiency of operations
over Zonotopes such as closure~\cite{ballou-2022-increm-trans,
chawdhary-2018-increm-closin-octag, gange-2021-fresh-look,
mine-2006-octag-abstr-domain, schwarz-2023-octag-revis}.

While new algorithms provide their complexity estimates, empirically evaluating
their runtimes remains crucial to comprehensively assessing their impact.
Commonly, such evaluations are performed in the context of a verifier and its
target, e.g., a data-flow analyzer using Zones~\cite{mine-2001-new-numer,
mine-2004-weakl-relat} over a set of programs.  However, depending on program
structure and semantics~\cite{brunner-2020-paclab}, one may or may not detect
the effect of the new operation over the abstract domain.  As such, the
question shifts to whether the set of programs are representative or the
implementation of the new algorithm is inefficient and requires additional
tuning.  Because of the complexity of Zonotope states, it is difficult to
assess whether a verifier produces states with properties that a novel
operation algorithm sufficiently takes advantage.

This problem is known to other research communities such as software
engineering and compiler optimization community, which they solve by
establishing microbenchmarking frameworks~\cite{laaber-2018-evaluat-open}.
Microbenchmarking isolates the effects of a specific technique such as a
certain optimization on syntactically generated code with desired features.  In
this work, we introduce \texttt{muRelBench}\footnote{Available on GitHub:
\url{https://github.com/fmsea/muRelBench}.}, an extensible microbenchmarking
framework for Zonotopes that is built on top of the
\texttt{JMH}~\cite{web-2024-jmh, web-2024-jmh-github} profiling tool for Java
programs.  \texttt{muRelBench} eliminates verifier and program dependencies and
focuses on specific operations of parameterized Zonotope states.

For a given type of Zonotope domain, \(Z\) and its operation \(ops\),
\texttt{muRelBench} takes as input a set of predefined parameters for each
characteristic of the corresponding \(Z\) typed abstract domain.  Then the
framework exhaustively generates abstract states corresponding to each element
of the Cartesian product of those parameters and applies \(ops\) and
correctness checks, if any, within the \texttt{JMH} context.  Upon the
completion of experiments, \texttt{muRelBench} writes the runtime results for
each abstract domain to a variety of output formats, including \gls{CSV} files
or \texttt{JSON} files, which researchers can use for further analysis and
evaluation.

In its current version, generation of abstract states is parameterized by the
number of variables and variable connectivity for the
Octagons(\(Z\))~\cite{mine-2006-octag-abstr-domain}.  Thus, synthetically
generated matrices that encode Octagon states vary in their size and variable
relation \emph{density}.  \texttt{muRelBench} implements two closure operations
(\(ops\)): Full Transitive Closure (using Floyd-Warshall all pairs shortest
path~\cite{cormen-2009-introd-algor}) and
Chawdhary~\cite{chawdhary-2018-increm-closin-octag} incremental closure.
However, as we describe in the next section, \texttt{muRelBench} can be easily
extended to different \(Z\) and \(ops\) types.

This microbenchmarking framework has the following three key features: (1)
dynamic generation of parameterized abstract states, (2) application of user
defined operations on them, and (3) checks to user-defined properties, e.g.,
pre/post conditions on Zonotope states before and after executing operations.
We believe that \texttt{muRelBench} will help rapid prototyping of abstract
operations and evaluating the efficiency of existing implementations.

In the next section~\ref{sec:framework}, we describe framework details and
explain how the framework generates different abstract states.  To demonstrate
the usefulness of \texttt{muRelBench}, in Section~\ref{sec:benchmarks}, we
present a case study on runtime data of two closure operators on Octagon
states.  We conclude the paper with future work on \texttt{muRelBench}.



\section{\texttt{muRelBench} Framework}\label{sec:framework}%

Figure~\ref{fig:uml} provides an overview of \texttt{muRelBench}'s components.
In the dashed, rounded rectangle are user-defined components of an abstract
domain type \(Z\), operations, e.g., \(ops1\), and property checks of the state
after \(ops1\) modifies the abstract state.  These bindings are defined at
compile-time.  A \emph{state generator} component takes generation parameters
\(N\) and \(D\)--- number of variables and density of the synthetic \gls{DBM},
respectively, and \(Z\) type, and randomly (up to the seed) generates \(N
\times D\) abstract states.

The \emph{Benchmarking} component takes the generated states and applies
\(ops1\) state operation and checks the results with \(check1\).  The component
also takes the runtime parameters for \texttt{JMH} that defines what type or
runtime data to collect and how many times to repeat the experiments.  Upon
completion, the data is written to the console and, optionally, to an output
file.

\begin{figure}
  \centering{}
  \definecolor{softblue}{HTML}{c3e5f5}
\definecolor{darkblue}{HTML}{3465a4}
\begin{tikzpicture}[%
  font=\large,
  connection/.style={color=darkblue, draw,-{Stealth[round]}},
  comp/.style={draw, rectangle, rounded corners, inner sep=0.5em, minimum size=1em},
  param/.style={font=\small,
    draw,
    trapezium,
    trapezium left angle=60,
    trapezium right angle=120,
    fill=gray!10}
  ]%

  \draw[fill=softblue, rounded corners, dashed] (-3.75, 4.5) rectangle (2.5, 3.5);
  \node[comp] at (-3, 4) (ops1) {\(ops1\)};
  \node[comp] at (-1, 4) (check1) {\(check1\)};
  \node at (1, 4) {\dots};
  \node[comp] at (2, 4) (Z) {\(Z\)};

  \node[comp, fill=softblue] at (2, 2) (gen) {\emph{State generator}};
  \node[param, inner sep=-0.4em, text width=2.5cm] at (6, 3.5) (cparams) {
    \begin{align*}
    N &= \{25,50,\cdots\},\\[0pt]
    D &= \{.1,.2,\cdots\}\\[0pt]
  \end{align*}};
  \node[param] at (-4.75, 0) (rparams) {\(R = \{\text{average}, 3 \text{ runs}\}\)};
  \node[comp, fill=softblue] at (0, 0) (bench) {\emph{Benchmarking}};
  \node[param] at (0, -2) (output) {Runtime data for each \(N \times D\)};

  \draw[connection] (rparams) edge (bench);
  \draw[connection] (cparams) -- (6, 2) -- (gen);
  \draw[connection] (Z) edge (gen);
  \draw[connection] (bench) edge (output);
  \draw[connection] (ops1) -- (-3, 1.5) -- (-1, 1.5) -- (-1, 0.35);
  \draw[connection] (check1) -- (-1, 2) -- (0, 2) -- (bench);
  \draw[connection] (gen) -- (2, 0) -- (bench);

\end{tikzpicture}

  \caption{Component diagram of \texttt{muRelBench}, specifying the framework's
    and user-defined components.}%
  \label{fig:uml}
\end{figure}

The framework is implemented in Java and uses interfaces and abstract classes
to provide extension points for user-defined components.  \texttt{JMH} provides
a strong foundation for constructing and executing profiling benchmarks whilst
minimizing confounding runtime variables such as \gls{JVM} startup, \gls{JIT}
warmup, and \gls{GC} pauses.  Specifically, \texttt{muRelBench} defaults to
\emph{three} warmup iterations before executing \emph{five} experimental
iterations for each benchmark.  This way, the code under bench has a chance to
\gls{JIT} compile.  We do not specifically tackle the notorious issue of CPU
boosting and other dynamic scaling policies that generally plague benchmarks.

The framework currently has extension for Octagon abstract domain, i.e., \(Z =
Octagon\).  The implementation encodes Octagon constraints, which are
constraints of the form: \(\pm{}x\pm{}y \le c\), where \(x, y \in V\) where
\(V\) is the set of program variables and \(c \in \mathbb{I}\), where
\(\mathbb{I}\) is one of \(\mathbb{R}\), \(\mathbb{Q}\), or \(\mathbb{Z}\).
Octagons are encoded as a 2-dimensional matrix, e.g.,
\gls{DBM}~\cite{dill-1990-timin-assum} in the
\texttt{OctagonDifferenceBoundedMatrix} class.

To extend operations over Octagons, users would provide extensions to
\texttt{OctagonDifferenceBoundedMatrix}, overriding various operations with
their implementation they wish to test.  Furthermore, users provide additional
instances of \texttt{*Bench}, e.g., \texttt{JoinBench}.  Similar to
\texttt{JUnit}~\cite{web-2024-junit}, the naming follows convention:
\texttt{muRelBench} automatically includes classes containing the
\texttt{Bench} suffix.

\paragraph{User extension beyond Octagons} It is reasonably straightforward
to extend \texttt{muRelBench} with additional abstract domains.  A user must
provide three additional classes: the abstract container type for the domain,
e.g., \texttt{ZoneDBM} to add Zones~\cite{mine-2001-new-numer}; a builder for
the new abstract type; and finally, a \emph{state} type which provides a
container for the different parameterization sets for \texttt{JMH}.



\section{Octagons and Closure Operation Case Study}\label{sec:benchmarks}%

\paragraph{Benchmark Set Up} We examine the benefits of \texttt{muRelBench} in
a case study.  The framework randomly generates Octagons, varying the
\emph{density} of relations between variables to create a continuum of
synthetic instances.  This density progression roughly correlates to the
different instances of Octagons from real programs.  That is, early in
analysis, variables have a tendency to have few relations as only few program
statements are explored.  In the middle of analysis, after exploring several
assignment statements, variables become tightly coupled with one another.
Finally, after several fixed point iterations and widening operations, islands
of connectivity emerge~\cite{gange-2016-exploit-spars, gange-2021-fresh-look,
singh-2015-makin-numer}.  Furthermore, we also vary the number of variables of
the synthetic Octagons to account for different programs sizes.

For this case study, we generate Octagons with \(25\), \(50\), and \(100\)
program variables, i.e., \(50\), \(100\), and \(200\) variables using the
Octagon variable encoding~\cite{mine-2006-octag-abstr-domain}.  For each size,
we generate Octagons with \(10\%-90\%\) density, in \(10\%\) increments.  The
Cartesian product of these parameters results in \(27\) Octagon instances.
Furthermore, while other tools such as \texttt{Apron}~\cite{jeannet-2009-apron}
can also generate random, synthetic Octagons, we make a point to only generate
\emph{consistent} synthetic Octagons.

Using \texttt{JMH}, we default to \(3\) ``warmup'' iterations and \(5\)
experimental iterations for each benchmark.  Thus, for a single benchmark, the
operation under test executes \(216\) times.  However, we do provide options
for the user to modify and otherwise specify their own desired warmup and
experimental iterations, among other options available via \texttt{JMH}.

\paragraph{Case Study}  In this case study, we chose to evaluate different
closure algorithms for Octagon abstract domain.  Closure represents a critical
operation for static program analysis and abstract interpretation because it
provides critical functions: normalization for equality comparisons for
\gls{DFA}~\cite{ballou-2022-increm-trans} and precision benefits for other
domain operations such as \gls{LUB}~\cite{mine-2004-weakl-relat}.

Canonicalizing or normalizing Octagon states is a necessary operation because
an octagonal bounded region can be represented by infinitely many different
Octagons.  The closure operation normalizes an Octagon by making explicit
implicit edges and minimizing edge weights between variables within the
Octagons.  In the simplest case, this amounts to computing the all-pairs
shortest-path problem for the directed, weighted graph used to represent the
Octagon.

There exist several algorithms for computing the all-pairs-shortest-path
problem for weighted-directed graphs such as Floyd-Warshall and Bellman-Ford
algorithms~\cite{cormen-2009-introd-algor}.  While these algorithms are
relatively simple and straightforward to implement, their cost can be
excessive.  Floyd-Warshall, for example has cubic time complexity,
\(\Theta(n^3)\), where \(n\) is the number of variables in the abstract Octagon
state.

Chawdhary et~al.~\cite{chawdhary-2018-increm-closin-octag} proposed an
incremental closure algorithm for Octagons which uses code motion and hoisting
to minimize the number of comparisons required to incrementally close an
Octagon.  Thus, they were able to reduce the incremental closure, a modified
Floyd-Warshall, to \(O(20n^2 - 4n)\).

\begin{table}
  \centering
  \begin{tabular}{llcc}
  \toprule
  \bf{Closure} & \bf{Program} & \bf{Mean (ms)} & \bf{\(\sigma\)} \\[0pt]
  \midrule{}
  \multirow{2}{*}{Floyd-Warshall} & Fibonacci & 144 & 32.2 \\[0pt]
               & Loop & 46.8 & 3.1 \\[0pt]
  \midrule{}
  \multirow{2}{*}{Chawdhary} & Fibonacci & 117 & 5.1 \\[0pt]
               & Loop & 49.6 & 10.3 \\[0pt]
  \bottomrule
\end{tabular}

  \caption{Small programs used to demonstrate performance characteristics of
    using different closure algorithms.}%
  \label{tab:closure-stats}
\end{table}

Clearly, these two algorithms should have a different runtime growth with the
increased number of variables.  We first examined their result in the context
of \gls{DFA} on two small programs to see if any differences can be detected.
Table~\ref{tab:closure-stats} shows the results of the full-closure algorithm
Floyd-Warshall and the Chawdhary et~al.'s incremental closure.  The data is
averaged over five executions and includes the mean runtime for each along with
their standard deviation, \(\sigma\).  As the data shows, the results are not
entirely conclusive since on the \texttt{Loop} program, Floyd-Warshall performed
better while Chawdhary runs faster on \texttt{Fibonacci}.  When we analyzed the
properties of the two programs, we discovered that \texttt{Fibonacci} algorithm
had a maximum of six variables with density of \(72\%\) and \texttt{Loop}
program had two variables with no density, which is purely interval.

\begin{figure}[t]
  \centering
  \input{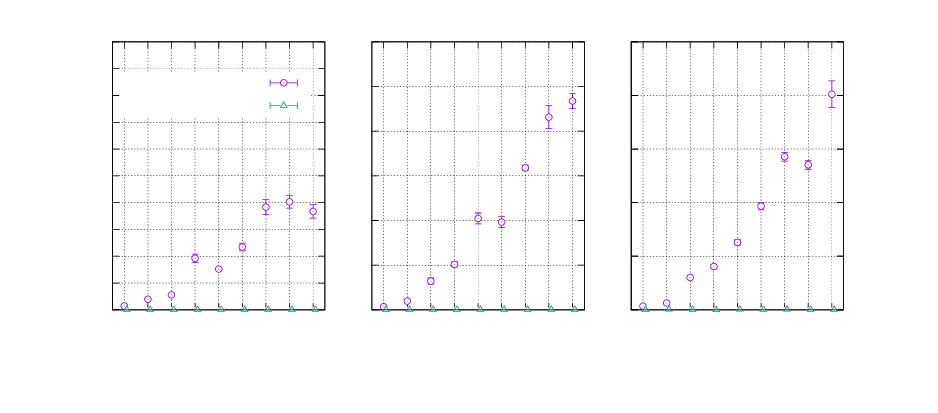}
  \caption{Plots of microbenchmark results of closure operations, each subplot
    varies the number of variables, each sample varies the connectivity of
    program variables.}%
  \label{fig:microbenchmark-results}
\end{figure}

Plots in Fig.~\ref{fig:microbenchmark-results} show the results of the
comparison of the two closure algorithm on benchmarks that \texttt{muRelBench}
generates and runs.  Each plot presents runtime data for different values of
\(N\) while varying in density of connections between variables.  Using this
detailed data, we can discern clear differences between the two algorithms
under comparison.  Specifically, when the density is small, in each variable
instance, the two algorithms seem to perform similarly.  However, as soon as
the density starts to climb above 30\%, the incremental algorithm of Chawdhary
et~al.\ clearly computes closure operations more efficiently than that of
Floyd-Warshall.  Furthermore, variable density shows a significant impact on
the runtime for Floyd-Warshall compared to Chawdhary et~al.'s, which remains
constant.

\begin{figure}[t]
  \centering
  \input{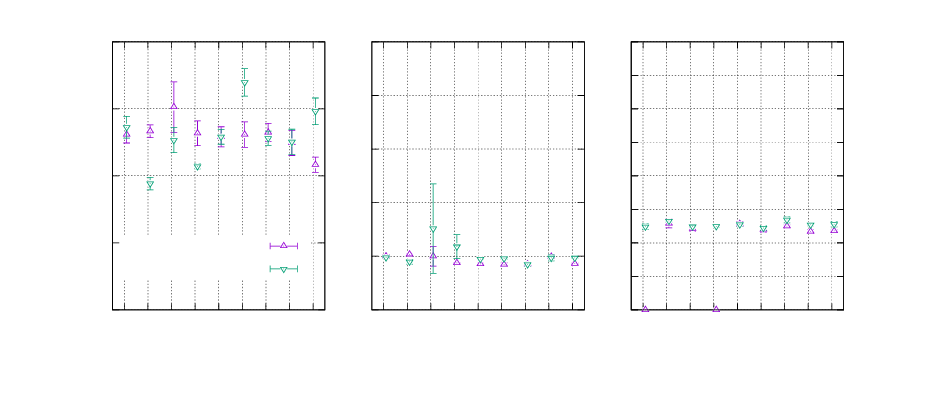}
  \caption{Plots of microbenchmark results for incremental closure operations.
Each subplot again varies by number of variables, each sample varies by
connectivity of the program variables.}%
  \label{fig:microbenchmark-results-incremental}
\end{figure}

While it is expected to see a vast performance gap between full closure and
incremental closure, we can zoom into the incremental approach and examine two
different incremental approaches to closure.
Figure~\ref{fig:microbenchmark-results-incremental} shows similar set of plots
between the Chawdhary et~al.\ incremental closure algorithm and an incremental
closure algorithm similar to the original proposed by
Min\'{e}~\cite{mine-2004-weakl-relat}.  Aside from some expected statistical
noise for the small variable size, \(N = 25\), these two closure algorithms
perform nearly identically.  Furthermore, density does not appear to
significantly contribute to the runtime of the incremental algorithms under
consideration, as was the case for full closure.

It may be tempting to use small programs to quickly validate new algorithm
performance, however, such small programs often do not demonstrate realized
benefit, as shown in Table~\ref{tab:closure-stats}.  The results shown in
Fig~\ref{fig:microbenchmark-results} more acutely capture the performance
differences between full closure and an incremental closure.  The results of
the benchmark would thus encourage further experimentation.  However, the
results from Fig~\ref{fig:microbenchmark-results-incremental} encourage further
algorithmic refinements before more intensive experimental study.  That is,
using a tool like \texttt{muRelBench} can save time and focus efforts by having
a smaller but appropriate set of benchmarks to validate algorithmic
improvements\footnote{The current set of benchmarks were run using GitHub
Actions on a default Ubuntu 24.04 Linux-2 instance, which currently only takes
approximately 30 minutes.}.



\section{Conclusion and Future Work}\label{sec:concl}%

In this paper, we present the \texttt{muRelBench} benchmarking framework to the
abstract interpretation research community.  This framework offers standardized and
uniform support for comparing various operations within Zonotope abstract
domains.  When developing new algorithms or new abstract domains, a standard
set of benchmarks and a common framework to easily test them helps convince the
community of their value.

Our framework of generated benchmarks invites many improvements and future work
to better situate it for the research community and software engineers at
large.  For example, we invite contributions of additional algorithms to be
added to the suite, so others can use the results in their comparisons.
Additionally, more parameters could provide a wider surface area of study for
different Zonotope operations.



\section*{Acknowledgments}

The work reported here was partially supported by the U.S. National Science
Foundation under award CCF-19-42044.


\bibliographystyle{splncs04}
\bibliography{bibfile, bibfile01}

\begin{thebibliography}{10}
\providecommand{\url}[1]{\texttt{#1}}
\providecommand{\urlprefix}{URL }
\providecommand{\doi}[1]{https://doi.org/#1}

\bibitem{ballou-2022-increm-trans}
Ballou, K., Sherman, E.: Incremental transitive closure for zonal abstract
  domain. In: Deshmukh, J.V., Havelund, K., Perez, I. (eds.) NASA Formal
  Methods. pp. 800--808. Springer International Publishing, Cham (5 2022).
  \doi{10.1007/978-3-031-06773-0_43},
  \url{http://dx.doi.org/10.1007/978-3-031-06773-0_43}

\bibitem{brunner-2020-paclab}
Brunner, R., Dyer, R., Paquin, M., Sherman, E.: Paclab: A program analysis
  collaboratory. In: Proceedings of the 28th ACM Joint Meeting on European
  Software Engineering Conference and Symposium on the Foundations of Software
  Engineering. ESEC/FSE ’20, ACM (11 2020). \doi{10.1145/3368089.3417936},
  \url{http://dx.doi.org/10.1145/3368089.3417936}

\bibitem{chawdhary-2018-increm-closin-octag}
Chawdhary, A., Robbins, E., King, A.: Incrementally closing octagons. Formal
  Methods in System Design  \textbf{54}(2),  232--277 (1 2018).
  \doi{10.1007/s10703-017-0314-7},
  \url{http://dx.doi.org/10.1007/s10703-017-0314-7}

\bibitem{web-2024-junit}
Contributors, M.: Junit (2024), \url{https://junit.org/}

\bibitem{cormen-2009-introd-algor}
Cormen, T., Leiserson, C., Rivest, R., Stein, C.: Introduction to Algorithms.
  Computer science, McGraw-Hill (2009). \doi{10.1.1.708.9446},
  \url{https://books.google.com/books?id=aefUBQAAQBAJ}

\bibitem{dill-1990-timin-assum}
Dill, D.L.: Timing assumptions and verification of finite-state concurrent
  systems. Lecture Notes in Computer Science pp. 197--212 (1990).
  \doi{10.1007/3-540-52148-8_17},
  \url{http://dx.doi.org/10.1007/3-540-52148-8_17}

\bibitem{gange-2021-fresh-look}
Gange, G., Ma, Z., Navas, J.A., Schachte, P., S{\o}ndergaard, H., Stuckey,
  P.J.: A fresh look at zones and octagons. ACM Transactions on Programming
  Languages and Systems  \textbf{43}(3),  1--51 (9 2021).
  \doi{10.1145/3457885}, \url{http://dx.doi.org/10.1145/3457885}

\bibitem{gange-2016-exploit-spars}
Gange, G., Navas, J.A., Schachte, P., S{\o}ndergaard, H., Stuckey, P.J.:
  Exploiting sparsity in difference-bound matrices. Lecture Notes in Computer
  Science pp. 189--211 (2016). \doi{10.1007/978-3-662-53413-7_10},
  \url{http://dx.doi.org/10.1007/978-3-662-53413-7_10}

\bibitem{ghorbal-2009-zonoto-abstr}
Ghorbal, K., Goubault, E., Putot, S.: The zonotope abstract domain taylor1+.
  Lecture Notes in Computer Science pp. 627--633 (2009).
  \doi{10.1007/978-3-642-02658-4_47},
  \url{http://dx.doi.org/10.1007/978-3-642-02658-4_47}

\bibitem{jeannet-2009-apron}
Jeannet, B., Min\'{e}, A.: Apron: A library of numerical abstract domains for
  static analysis. Lecture Notes in Computer Science pp. 661--667 (2009).
  \doi{10.1007/978-3-642-02658-4_52},
  \url{http://dx.doi.org/10.1007/978-3-642-02658-4_52}

\bibitem{jordan-2022-zontope-neural}
Jordan, M., Hayase, J., Dimakis, A., Oh, S.: Zonotope domains for lagrangian
  neural network verification. Advances in Neural Information Processing
  Systems  \textbf{35},  8400--8413 (2022)

\bibitem{laaber-2018-evaluat-open}
Laaber, C., Leitner, P.: An evaluation of open-source software microbenchmark
  suites for continuous performance assessment. In: Proceedings of the 15th
  International Conference on Mining Software Repositories. ICSE ’18, ACM (5
  2018). \doi{10.1145/3196398.3196407},
  \url{http://dx.doi.org/10.1145/3196398.3196407}

\bibitem{mine-2001-new-numer}
Min\'{e}, A.: A new numerical abstract domain based on difference-bound
  matrices. Lecture Notes in Computer Science pp. 155--172 (2001).
  \doi{10.1007/3-540-44978-7_10},
  \url{http://dx.doi.org/10.1007/3-540-44978-7_10}

\bibitem{mine-2004-weakl-relat}
Min\'{e}, A.: {Weakly Relational Numerical Abstract Domains} (12 2004),
  \url{https://pastel.archives-ouvertes.fr/tel-00136630}

\bibitem{mine-2006-octag-abstr-domain}
Min\'{e}, A.: The octagon abstract domain. Higher-Order and Symbolic
  Computation  \textbf{19}(1),  31--100 (3 2006).
  \doi{10.1007/s10990-006-8609-1},
  \url{http://dx.doi.org/10.1007/s10990-006-8609-1}

\bibitem{web-2024-jmh}
Oracle: jmh (2024), \url{https://openjdk.org/projects/code-tools/jmh/}

\bibitem{web-2024-jmh-github}
Oracle: openjdk/jmh (2024), \url{https://github.com/openjdk/jmh}

\bibitem{schwarz-2023-octag-revis}
Schwarz, M., Seidl, H.: Octagons revisited. Lecture Notes in Computer Science
  p. 485–507 (2023). \doi{10.1007/978-3-031-44245-2_21},
  \url{http://dx.doi.org/10.1007/978-3-031-44245-2_21}

\bibitem{singh-2015-makin-numer}
Singh, G., P\"{u}schel, M., Vechev, M.: Making numerical program analysis fast.
  ACM SIGPLAN Notices  \textbf{50}(6),  303--313 (8 2015).
  \doi{10.1145/2813885.2738000},
  \url{http://dx.doi.org/10.1145/2813885.2738000}

\end{thebibliography}

\end{document}